# Evidence for a robust sign-changing s-wave order parameter in monolayer films of superconducting Fe (Se,Te)/Bi$_2$Te$_3$

Guannan Chen[1]*, Anuva Aishwarya[1]*, Mark R. Hirsbrunner[1], Jorge Olivares Rodriguez[1], Lin Jiao[1], Lianyang Dong[2], Nadya Mason[1], Dale Van Harlingen[1], John Harter[2], Stephen Wilson[2], Taylor L. Hughes[1], Vidya Madhavan[1]

[1]*Department of Physics and Materials Research Laboratory, University of Illinois Urbana-Champaign, Urbana, Illinois 61801, USA*

[2]*Materials department, University of California, Santa Barbara, USA*

*equal contribution

**Abstract:**

The Fe-based superconductor Fe (Se,Te) combines non-trivial topology with unconventional superconductivity and may be an ideal platform to realize exotic states such as high-order topological corner modes and Majorana modes. Thin films of Fe (Se,Te) are important for device fabrication, phase sensitive transport measurements and for realizing proposals to engineer higher-order modes. However, while bulk Fe (Se,Te) has been extensively studied with a variety of techniques, the nature of the superconducting order parameter in the monolayer limit has not yet been explored. In this work, we study monolayer films of Fe (Se,Te) on Bi$_2$Te$_3$ with scanning tunneling spectroscopy and Bogoliubov quasiparticle interference (BQPI). We discover that the monolayer Fe (Se,Te)/Bi$_2$Te$_3$ heterostructures host a robust, multigap superconducting state that strongly resembles the bulk. BQPI maps at the gap energies show a strong spatial modulation, oriented 45 degrees to the Fe-Se bond direction. Analysis of the phase-referenced quasiparticle interference signal reveals a sign-changing s-wave order parameter similar to the bulk. Moreover, we observe a unique pattern of sign changes in the BQPI signal which have not been observed in the bulk. Our work establishes monolayer Fe (Se,Te)/Bi$_2$Te$_3$ as a robust multi-band unconventional superconductor and sets the stage for explorations of non-trivial topology in this highly-tunable system.

Iron-based superconductors (Fe-Sc) are a primary target in the modern hunt for topological boundary modes and non-abelian anyons. Fe-Sc exhibits a high transition temperature ($T_c$) unconventional superconducting state that occurs near a magnetically ordered state, with antiferromagnetic spin-fluctuations proposed as a promising candidate for the mediator of superconductivity [1]. The layered iron-chalcogenide Fe-Sc Fe (Se, Te) has been proposed to be topologically non-trivial and has a large body of experimental evidence for Majorana bound states in vortex cores [2-5] and near a variety of different defects [6-10]. Interestingly, recent theory and experiments indicate that a monolayer of Fe (Se, Te) also possesses a band-inversion [11,12], and theory has predicted a tunable and scalable Majorana platform with a promise of high-order topological superconductivity in the same [13]. Growing high quality Fe (Se, Te) films and measuring their properties is an important first step in using these materials for quantum applications. It is therefore crucial to understand the electronic structure and the nature of the superconducting order-parameter in monolayer Fe (Se, Te). In this work, we employ sub-kelvin scanning tunneling microscopy/spectroscopy (STM/S) imaging to study Bogoliubov quasiparticle interference (BQPI) in the superconducting state in monolayer Fe (Se,Te) on $Bi_2Te_3$. Our experimental results combined with theoretical modelling and numerical simulations confirm a sign-changing s-wave order parameter (OP) for the monolayer and provide key insights into the unusual nature of impurity scattering originating in the substrate.

Monolayer and bilayer films of Fe (Se,Te) were grown using molecular beam epitaxy (MBE) on cleaved single crystals of $Bi_2Te_3$. Previous studies have shown that $Bi_2Te_3$ is an ideal substrate for Fe (Se,Te) thin films [14]. This is because by a magical coincidence the in-plane lattice constants of $Bi_2Te_3$ (~4.2 Å) and Fe (Se,Te) (average value ~3.79 Å) are almost perfectly at the ratio of $2:\sqrt{3}$. With this ratio, the two materials with 4- and 6-fold in-plane symmetries make a perfect lattice match along the b-axis [15]. After the MBE growth, the films were transferred from the MBE chamber to the STM via vacuum suitcase without exposure to the ambient environment.

A typical large-scale STM topographic image of this heterostructure shown in Fig. 1a displays Fe (Se,Te) islands as well as the underlying $Bi_2Te_3$ surface. An atomically resolved topography of $Bi_2Te_3$ exhibiting a triangular lattice with a lattice constant of 4.2 Å is shown in Fig. 1b. In contrast, the topography taken on Fe (Se,Te) islands (Fig. 1c) shows a square lattice of the top chalcogen atom layer with a lattice constant of 3.8 Å. The brighter sites correspond to Te atoms as their electronic orbitals are more extensive than the Se atoms' [16-18]. A statistical count of the bright and dark sites in the topography indicates a Se:Te ratio of ~2:3 (i.e. 0.4:0.6). This ratio of Se:Te is deep in the topologically non-trivial regime of the Fe (Se,Te) phase diagram as predicted by theoretical calculations [11]. It is worth noting that substrate effects have been known to alter superconducting properties and critical Te concentration of topological-trivial phase transition in the Fe-Scs. In this work we focus on the superconducting order parameter and further studies are needed to unequivocally establish the topological nature of our heterostructure.

There are three different types of Fe (Se,Te)/$Bi_2Te_3$ heterostructures present in our samples (see Supplementary Figure 1 and 2); here we primarily study the half-embedded Fe (Se,Te) terraces as they have the most robust superconductivity (refer to the Supplement for a discussion of superconductivity in the other heterostructures) and clear interference patterns. Half-embedded terraces consist of a single trilayer stack (Se/Te) –(Fe) –(Se/Te) partly embedded into the $Bi_2Te_3$ surface, as shown in Fig. 1f [14]. Fig. 1e shows the height profile obtained along the green line in Fig. 1a. The height profile shows two islands on either side of the central $Bi_2Te_3$ region which are ~0.26 nm higher than $Bi_2Te_3$ surface. The lattice constant along the *c*-axis of Fe (Se,Te) ranges from 0.55 to 0.62nm (depending on the concentration of Se) [19], so the height difference of 0.26 nm is slightly less than half of the *c*-axis lattice constant and is consistent with a half-embedded Fe (Se,Te) terrace.

Before proceeding further, it is useful to describe the band structure of Fe (Se,Te). The real-space unit cell of Fe (Se,Te) contains two Fe atoms since the alternating chalcogen atoms are inequivalent. The inequivalence of the local environment around the two Fe atoms is often ignored and a smaller unit cell containing one Fe atom is used. The Brillouin zone (BZ) of the 1-Fe unit cell is indicated by the dashed line in Fig. 1d, and the smaller "folded" BZ of the 2-Fe unit cell is shown by the solid line. Throughout this work we employ the 2-Fe unit cell BZ, except when performing numerical calculations. The inequivalence of the Fe atoms does not impact our results, so we take advantage of the smaller number of bands that appear in the 1-Fe unit cell in our simulations. In the 2-Fe BZ, the band structure consists of two hole-pockets around the $\Gamma$ point (labelled $\alpha$ and $\beta$) and two electron-pockets around the M point (both labelled $\gamma$).

To characterize the superconducting state of the half-embedded Fe (Se,Te) islands, we measure a dense grid of dI/dV spectra over a 30 x 30 nm area as shown in Fig. 2a. Spectra along the white lines in Fig. 2a and b are plotted in Fig. 2c. At each spatial location we see clear s-wave gaps with multiple coherence peaks as observed previously in bulk samples [3,6,9, Supplementary Figure 3]. These correspond to gaps on the different electron and hole pockets in Fe (Se,Te). As was done previously for the bulk samples, the we characterize the gap magnitude using an algorithm developed to identify the pairs of coherence peaks for each dI/dV spectrum [9]. Using this algorithm, we determine the gap magnitudes at each pixel in the dI/dV map. Compiling the results, we obtain the map presented in Fig. 2b, where the colors represent the different number of gaps found at each pixel. A statistical analysis of the gap magnitude in this area reveals a clear multimodal distribution, as summarized in the histogram shown on Fig. 2c. Performing a gaussian fit produces mean gap values of 1.2±0.1, 1.6 ±0.2, and 2.2±0.25 meV. The next step is to assign these gaps to the Fermi surface pockets.

We start with the generic Fe (Se,Te) family Fermi surface which has two hole-pockets at the $\Gamma$ point (labelled $\alpha$ and $\beta$) and two electron-pockets at the M-point (both labelled $\gamma$) as shown in Fig. 1d. Typically the two hole-pockets and the electron pockets all have different gap magnitudes. Experimental data have established that bulk $FeSe_{0.45}Te_{0.55}$ has a mean value of 1.4 meV for the $\alpha$-band gap and 2.4 meV for the $\beta$-band gap [5, 9, 20-21]. Given that the bulk bands in Fe (Se,Te) are known to be highly two-dimensional [22-23] and that the gap structure on the films closely resembles the bulk (Supplementary Figure 3), we tentatively associate the coherence peaks in our data at 1.2 meV and 2.2 meV with the gaps on hole-like bands i.e., $\Delta_{\alpha'}$ = 1.2 meV and $\Delta_{\beta'}$ = 2.2 meV (the bands for our monolayer film will henceforth be referred to using this primed nomenclature). This assignment preempts our BQPI data as we show next, which confirms the existence of pockets at the $\Gamma$ and M points in the monolayer.

Our next step is to investigate the nature of the superconducting order parameter using phase referenced quasiparticle interference (PR-QPI). It is known that the Bogoliubov quasiparticles in a superconductor are scattered by impurities and interfere with each other giving rise to spatial oscillations in the local density of states (LDOS). This process is dubbed Bogoliubov quasiparticle interference (BQPI). It has been shown that phase-sensitive measurements of BQPI at the energy of the coherence peaks can provide momentum-resolved information about the complex OP of the superconducting state and thus provide clues to the pairing mechanism [24-26]. Specifically, theoretical and experimental studies have demonstrated that the relative phase of the superconducting OP on different pockets in the BZ can be determined by STM-based phase-referenced QPI (PR-QPI) which has been used to study sign changes in the superconducting OP of both cuprate and iron-based superconductors [16,27-29]. Previous bulk studies of Fe (Se,Te) have shown that the relevant BQPI scattering wave-vectors are those that connect the hole pockets at $\Gamma$ to the electron pockets at M [16,27,30]. This interference process produces modulations in the real space LDOS with scattering wave-vector $q_{2aFe}=2\pi/a_{Fe}$, where $a_{Fe}$ is the Fe-Fe distance. PR-QPI relies on comparing the phase of these LDOS modulations in Fourier space at the energies of the coherence peaks associated with each band [30]. Calculations for bulk Fe (Se,Te) have shown for example that a sign change in the OP on different bands is reflected as a relative sign change in the PR-QPI signal between the coherence peaks above and below the Fermi energy.

Understanding the nature of the sources of scattering is very important for interpreting QPI experiments. The scattering sources are ideally point-like and well-separated from each other. However, our surfaces are very clean and topographic images do not reveal any impurities that could act as scattering centers. It is conceivable that scattering from terraces is responsible for our QPI signals, but such scattering typically produces an extended standing wave pattern that decays in intensity as a function of distance from the edge of the terrace. We observe no such thing, so we must instead conclude that the scattering centers are likely anti-sites or vacancies below the Fe (Se, Te) monolayer in the $Bi_2Te_3$ substrate [31]. This conclusion is further supported by the absence of any QPI signal in our bilayer systems (a monolayer of Fe (Se, Te) on top of a fully-embedded Fe (Se, Te) monolayer).

To visualize the BQPI for our films, we measure the differential conductance as a function of energy $g(r,E) \equiv dI/dV(r, E = eV)$ on a 150 x 150 grid on a 10 nm x 10 nm area. $dI/dV$ maps at the coherence peak energies $\pm E_{\alpha'}$ and $\pm E_{\beta'}$ associated with $\Delta_{\alpha'}$ and $\Delta_{\beta'}$ respectively, are shown in Fig. 3c-f (see Fig. 3a for the topography obtained simultaneously as the dI/dV maps). The data show modulations that have a periodicity of $2a_{Fe}$. Interestingly, we find that the phase of the modulations shows a relative shift at different energies. This is better seen in the high-resolution 5 nm x 5 nm area maps shown in the insets. The modulations at $g(r, +E_{\alpha'})$ and $g(r, -E_{\alpha'})$ (Fig. 3c and e) are phase reversed, as are the modulations at $g(r, +E_{\beta'})$ and $g(r, -E_{\beta'})$ (Fig. 3d and f).

To translate the real-space LDOS signatures into sign-changes of the OP across the electron/hole pockets, a momentum-space based machinery has been developed for PR-QPI. It is based on the simple idea that the Fourier transform (FT) (i.e., $g(q,E)$), of the real space QPI signal $g(r,E)$ is a complex number comprising of the amplitude and the phase (i.e., $|g(q,E)|$ and $e^{i\theta_{q,E}}$ respectively). The phase alone is arbitrary, but the relative phase between $g(q,E_1)$ and $g(q,E_2)$ has physical consequences. For example, the phase function $Re[e^{i(\theta_{q,E_1}-\theta_{q,E_2})}]$ will change sign between $E_1$ and $E_2$ if there is a contrast reversal in the real space LDOS modulation. We define the PR-QPI signal (or relative phase-amplitude) by $g_{PR}(q,E_1,E_2) \equiv |g(q,E_2)| Re\left[e^{i(\theta_{q,E_1}-\theta_{q,E_2})}\right] = |g(q,E_2)| \cos(\theta_{q,E_1} - \theta_{q,E_2})$. Tracking the relative phase amplitude of the superconducting OP as a function of energy in the FFTs provides information about all the sign changes of the OP across the several hole/electron pockets on the Fermi surface separated by wave-vector $q$ [30].

Fig. 4 shows the results of the PR-QPI analysis. This analysis was carried out on a large 30 nm x 30 nm area for better statistical sampling and all the phases were calculated relative to the phase of the signal in the *dI/dV* map at $E_{\alpha\prime}$, *i.e.*, $g(q,+E_{\alpha\prime})$. Following the PR-QPI protocol described above, we observe that the phase amplitude in the PR-QPI signal for the scattering vector q$_{2aFe}$ (averaged over a 4-pixel window) exhibits multiple sign-changes (Fig. 4b). Our first finding is that the PR-QPI signals at opposite sides of the Fermi energy are out of phase, i.e., $+E_{\alpha\prime}$ is out of phase from the signal at $-E_{\alpha\prime}$, and likewise at $+E_{\beta\prime}$ and $-E_{\beta\prime}$, as shown in Fig. 4b. Sign-changing PR-QPI signals of this nature have been previously reported in bulk Fe (Se,Te) and are evidence of a sign change of the OP between the hole ($\alpha\prime$ and $\beta\prime$) and electron ($\gamma\prime$) pockets [16,27]. The origin of the phase shift is believed to be an antiferromagnetic wave vector Q = (0, π) that mediates the Fermi surface nesting and eventually induces superconductivity. Our data thus shows that the half-embedded layer of Fe (Se,Te) on Bi$_2$Te$_3$ is a two-dimensional superconductor with a sign-changing s-wave OP.

Intriguingly, we also observe an additional relative sign change between the α′ and β′ coherence peaks on the same side of the E$_F$. This has not been observed experimentally or studied previously. This is seen as a contrast reversal in the pairs of real-space maps on the same side of the Fermi energy i.e., $g(r,+E_{\alpha\prime})$ and $g(r,+E_{\beta\prime})$ (Fig. 3c and d), or $g(r,-E_{\alpha\prime})$ and $g(r,-E_{\beta\prime})$ (Fig. 3e and f) respectively. The same phase shift can also be clearly detected when we compare the sign of the PR-QPI signal at q$_{2aFe}$ (Fig. 4b and Fig. 4d). This raises the question: is the additional phase-shift an indicator of a specific type of sign-changing s-wave OP, namely, s±, odd-parity s±, or orbital-antiphase s± [32-36]. To answer this question we perform exhaustive PR-QPI simulations using a detailed five-orbital model for Fe(Se, Te) and interpret the results according to the Hirschfeld, Altenfeld, Eremin, and Mazin (HAEM) approach to PR-QPI. The HAEM approach studies the energy-antisymmetrized QPI signal, $\delta\rho_-(E) = \rho(E) - \rho(-E)$, where $\rho(+E) = g(q,+E)$ (the experimental $\rho_-(E)$ has been plotted in Supplementary Figure 4). For a pair of Fermi pockets with coherence peaks at $E_1$ and $E_2$, the HAEM theory states that $\delta\rho(E)$ evaluated at the wavevector connecting the two pockets will be small and change signs between $E_1$ and $E_2$ if the order parameter is the same sign on both pockets. If instead the order parameter is of opposite sign on each pocket, $\delta\rho(E)$ will be large and not change signs in the same energy range. Our simulations are discussed in detail in the supplement. We consider three different s-wave sign changing OPs: s±, odd-parity s±, and orbital-antiphase s±. Our first finding is that all three OPs produce peaks in $\delta\rho(E)$ that correspond to the sign-changing PR-QPI peaks at opposite sides of the Fermi energy i.e., at $\pm E_{\alpha\prime}$ and $\pm E_{\beta\prime}$.

We find, however, that the relative sign of the QPI signals at $+E_{\alpha\prime}$ and $+E_{\beta\prime}$ (or $-E_{\alpha\prime}$ and $-E_{\beta\prime}$), on the same side of the Fermi energy are uncorrelated with the specific choice of sign-changing s-wave OP. This is in accordance with HAEM theory, which relates the magnitude of $\delta\rho(E)$ to the presence of an OP sign change, and the overall sign of $\delta\rho(E)$ to the sign of the relevant scattering matrix element. In a nutshell, while the HAEM analysis for a sign changing order parameter reproduces the experimentally observed sign change at opposite sides of $E_F$, it does not shed light on the sign change between $+E_{\alpha\prime}$ and $+E_{\beta\prime}$ (or $-E_{\alpha\prime}$ and $-E_{\beta\prime}$). To explain this additional phase shift, we study the orbital nature of the scattering processes contributing to the QPI in Fe (Se, Te). We find that the most plausible explanation of the additional PR-QPI sign change is that the in-plane Fe orbitals ($d_{xy}$ and $d_{x2-y2}$) and the out-of-plane Fe orbitals ($d_{xz}$, $d_{yz}$, and $d_{3z2-r2}$) that constitute the Fermi surface have oppositely-signed scattering matrix elements (See Supplementary Figure 8). The details of the simulations and the interpretation of the additional sign-change in terms of scattering potentials are provided in the supplement. Also included in the supplement is a plot (Supplementary Figure 5) of the experimental $\rho_-(E)$ for the wave vector connecting the electron pockets at the M points.

In conclusion, we have successfully grown superconducting films of Fe (Se,Te)/$Bi_2Te_3$ and have shown that they have well-defined s-wave gaps with multiple coherence peaks similar to the bulk. We note here that the observation of hard gaps similar to the bulk material is in stark contrast to a previous study of Fe (Se,Te) grown on BiSeTe [37] where V-shaped gaps and QPI similar to FeSe were observed. These differences between the two studies may be attributed to substrate induced doping and/or strain. We have proven using PR-QPI that the superconducting state is sign-changing s-wave in nature. Experimentally, our work opens the possibility of realizing Majorana modes or high-order corner modes as predicted to exist in this monolayer superconductor [13]. We additionally observe a systematic phase-shift on the same side of the Fermi energy, which has not been previously probed, that our simulations suggest might originate from= the orbital character of the different bands.

**Methods**

Fe (Se,Te) thin films were grown on $Bi_2Te_3$ using a home-built MBE UHV system with a base pressure $< 1 \times 10^{-9}$ Torr. The $Bi_2Te_3$ single crystal was adhered to the sample plate by carbon paste with a rod glued perpendicularly on top with silver epoxy. To obtain a fresh substrate surface, the $Bi_2Te_3$ single crystal was cleaved in the load lock of the MBE system under $\sim 5 \times 10^{-9}$ Torr environment. It was transferred into the MBE main chamber immediately afterwards for deposition. The substrate was heated up to 300 ℃ for an hour to degas the carbon paste and then cooled to room temperature. High purity Fe (99.95%) was evaporated from a high temperature Knudsen cell while the substrate stayed at room temperature. About 1 monolayer of Fe was deposited, where the flux of Fe was measured by quartz crystal microbalance. The sample was annealed post deposition at 300 ℃ for 30 mins.

After growth, the samples were transferred to an ultra-low temperature scanning tunneling microscope (STM) using a home-built "vacuum suitcase" with pressure less than $5 \times 10^{-10}$ Torr to prevent the degradation of the sample quality. The STM/S data were measured at 300 mK except the temperature dependent experiments. In the STM measurements, electrochemically etched and vacuum annealed tungsten tips were used.

**Acknowledgements:** The STM work at the University of Illinois was supported by DOE under Grant- DE-SC0022101. D.V.H, N.M, T.L.H. and M.H. were supported by the DOE "Quantum Sensing and Quantum Materials" Energy Frontier Research Center under Grant DE-SC0021238. Thin film growth was supported by the DOE "Quantum Sensing and Quantum Materials" Energy Frontier Research Center under Grant DE-SC0021238 and the Gordon and Betty Moore foundation EPiQS grant #9465 for instrumentation. V.M is a CIFAR Fellow in the Quantum Materials Program and acknowledges CIFAR for support. S.D.W., J.H., and L.D. gratefully acknowledge support via the UC Santa Barbara NSF Quantum Foundry funded via the Q-AMASE-i program under award DMR-1906325. This work made use of the Illinois Campus Cluster, a computing resource that is operated by the Illinois Campus Cluster Program (ICCP) in conjunction with the National Center for Supercomputing Applications (NCSA) and which is supported by funds from the University of Illinois at Urbana-Champaign.

**Author contributions:** G.C. and A.A. contributed equally and substantially to this work. G.C. and V.M. conceived the experiments. G.C., A.A., L.J. and J.O.R. obtained the STM data. L.D., J. H. and S.W. provided the single crystals of $Bi_2Te_3$. N.M. and D.V.H. provided input on interpreting the data. M.R.H. and T.H. carried out the theoretical modelling. A.A., G.C. and J.O.R. performed the analysis. A.A. M.R.H. and V.M. wrote the manuscript with input from all the authors.

**Competing interests:** The authors declare no competing interests.

**Data availability:** The experimental data will be available upon reasonable request.

**Figure 1**

**Figure 1 | Overview of Fe (Se,Te) lattice. a**, Topographic image of Fe (Se,Te) thin film grown on $Bi_2Te_3$ single crystal in a 100 nm × 100 nm area (bias voltage $V_b$ = 300 mV, tunneling current $I_t$ = 40 pA). Scale bar is 20 nm. **b**, **c**, Atomic resolution topographies of $Bi_2Te_3$ surface (Scale bar is 2 nm) and half-embedded Fe (Se,Te) island (Scale bar is 2 nm) respectively, taken within the orange and pink frames (8 nm × 8 nm) shown in **a**. **d**, Fermi surfaces of Fe (Se,Te) in the crystallographic BZ. The high-symmetry point labels and solid lines refer to the 2-Fe BZ, while the "unfolded" 1-Fe BZ is denoted by the dashed line. **e**, Height profile taken along the green line in **a**, which shows the height of half-embedded Fe (Se,Te) island is about 2.6 Å. **f**, Schematic lattice structure of the heterostructure between $Bi_2Te_3$ single crystal and half-embedded Fe (Se,Te) thin film.

**Figure 2**

**Figure 2 | Statistical analysis of superconducting gaps on a half-imbedded Fe(Se,Te) island. a**, Topography in a 30 nm × 30 nm field of view on Fe(Se,Te), where a dI/dV map was taken at 300 mK. Scale bar is 10 nm. **b**, Gap map, in the same area as **a**, describing the spatial distribution of the quantity of superconducting gaps. Orange, yellow, light green, and dark green colors indicate no gap, one gap, two gaps, or three gaps were found at each pixel, respectively. The gap values at each pixel were obtained through a multipeak-finding algorithm. **c**, Waterfall plot of dI/dV spectra obtained along the white line from the orange dot (top) to purple dot (bottom) as shown in **b**. **d**, Histogram of the gap values in the overall region which clearly shows three peaks with the mean gap values as 1.2, 1.6, and 2.2 meV.

**Figure 3**

**Figure 3 | Multiple contrast reversal in the real space of the Fe sublattice at various energies associated with the peaks of the superconducting gap. a,** Topography of a 10 nm x 10 nm area obtained on a monolayer of Fe(Se,Te) on $Bi_2Te_3$ which shows the Se/Te layer ($V_{Bias}$ = 10 mV, $I_t$ = 100 pA, and the scale bar is 1 nm long). **b,** Average dI/dV spectrum obtained at 300 mK in this area showing clear peaks associated with the $\alpha'$ and $\beta'$ bands. The peak energies that display modulations of the same phase have been highlighted with the same color. **c-f,** LDOS(r,E) maps (10 nm x 10 nm) obtained on the area shown in **a** showing the real space modulations different from the Se/Te layer associated with the underlying Fe sublattice. The black arrows in all the DOS maps are aligned with the crests of the modulation in the g(r, $+E_{\alpha'}$) slice. They fall *on* the crests of the modulation in the DOS maps of $-E_{\beta'}$ but are *off* for DOS maps $+E_{\beta'}$ and $-E_{\alpha'}$. Insets show maps at higher spatial resolution of the area around the black arrows. The white square and the translucent arrows are a guide to the eye. The arrows are *on* the modulations for **c** and **f** and *off* for **d** and **e**.

**Figure 4**

**Figure 4 | PR-QPI to show s± pairing and the sign-changes of the order parameter across $\alpha'$ and $\beta'$ bands.** (Sign change is indicated by blue or red colors). **a,** Fermi surface of an s± superconducting state of Fe (Se,Te) based on the 2-Fe unit cell model. The dashed line indicates the Brillouin Zone in the 1-Fe unit cell model. The q vector of the scattering process observed in our data is labelled as $q_{2aFe}$. The red and blue colors indicate the relative sign of the superconducting order parameter. **b**, Plot of the amplitude of the phase (relative to $+E\alpha'$ i.e. $\alpha'$ band) as a function of energy of the scattering vector $q_{2aFe}$ (summed inside the dotted region shown in c) in the phase-reference Fourier transforms of the DOS maps obtained on a 30 nm x

30 nm area. **c-e**, PR-QPI of the scattering vector $q_{2aFe}$ that connects the $\alpha'$, $\beta'$ and $\gamma'$ bands between various energies.

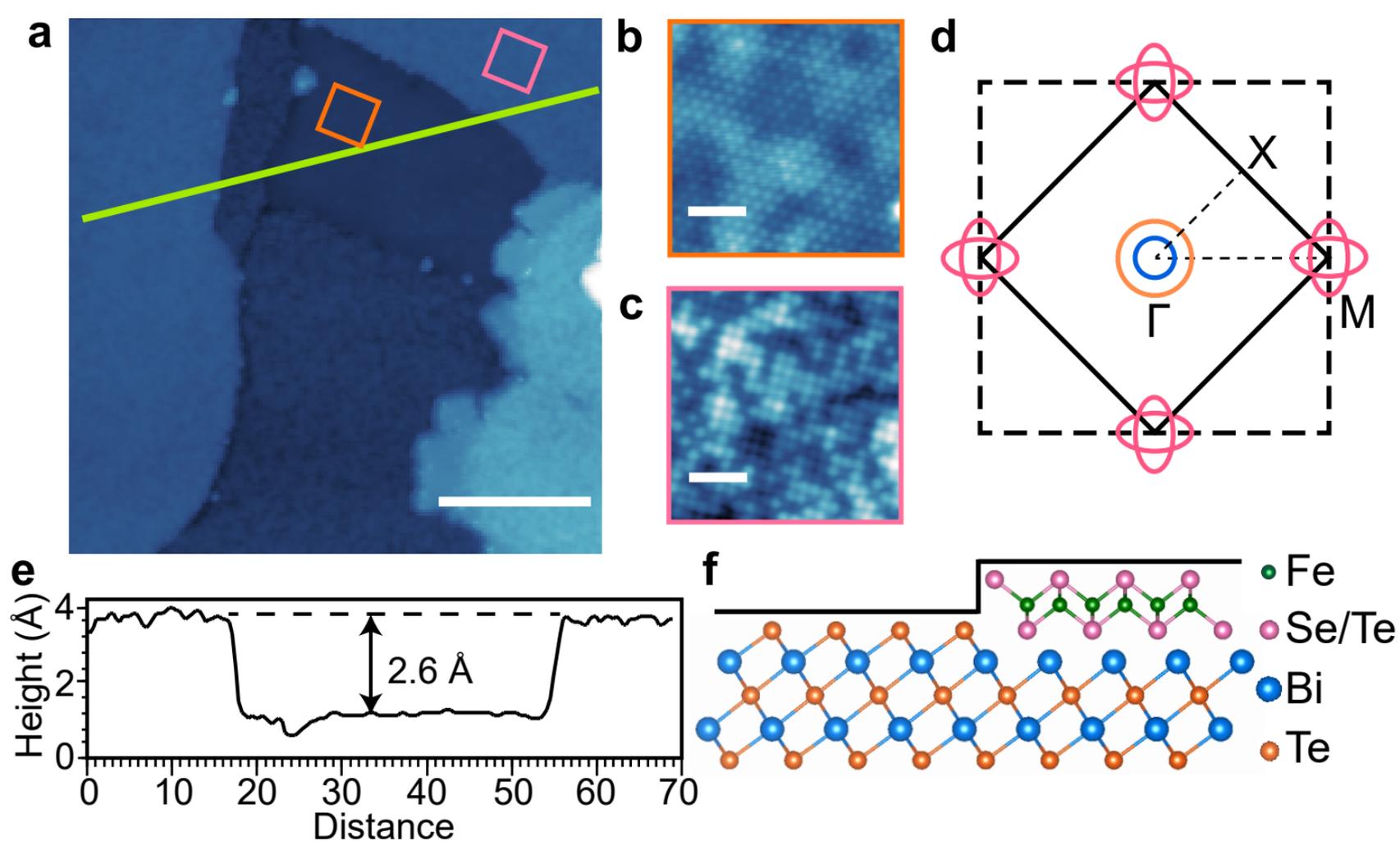

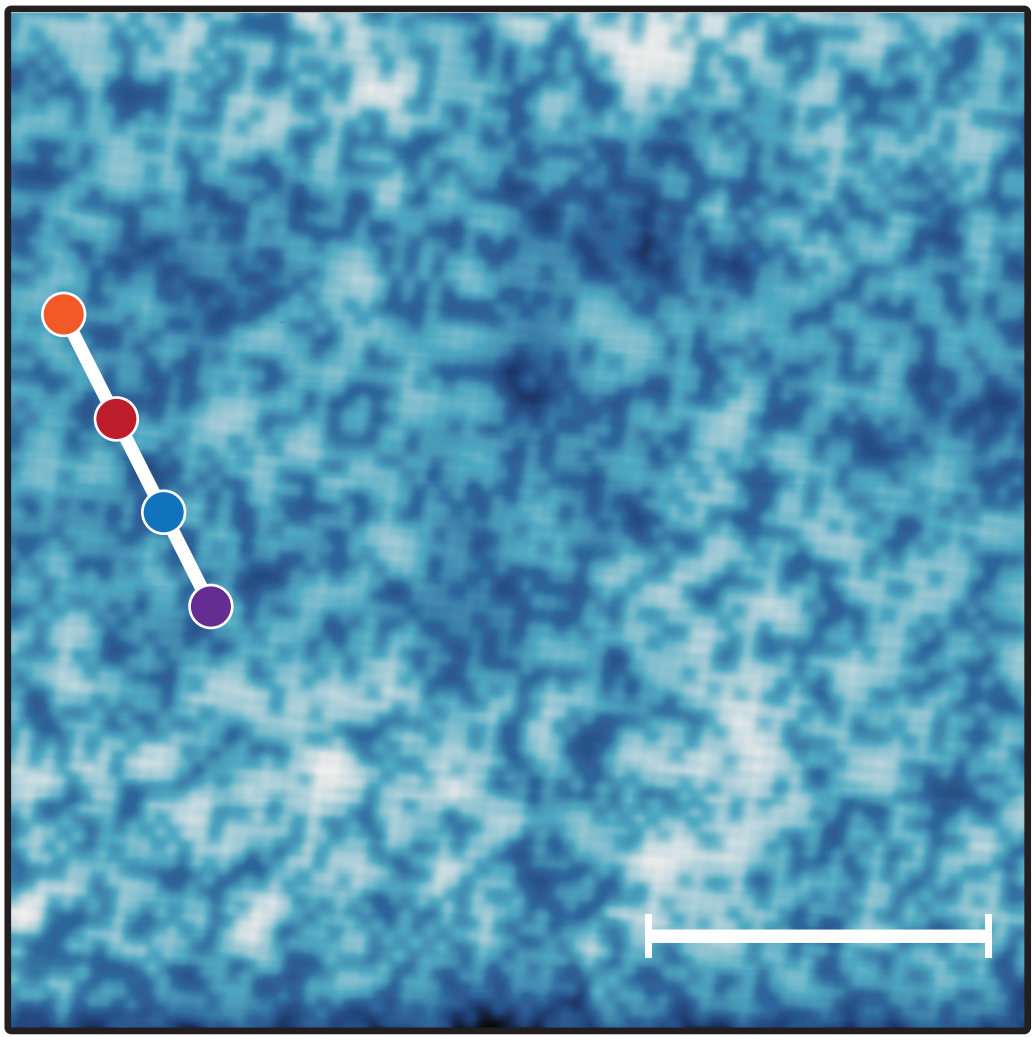
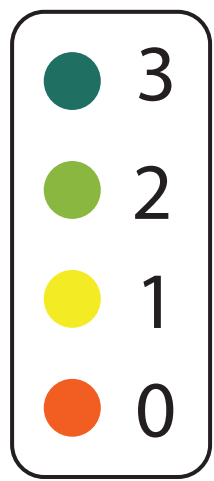
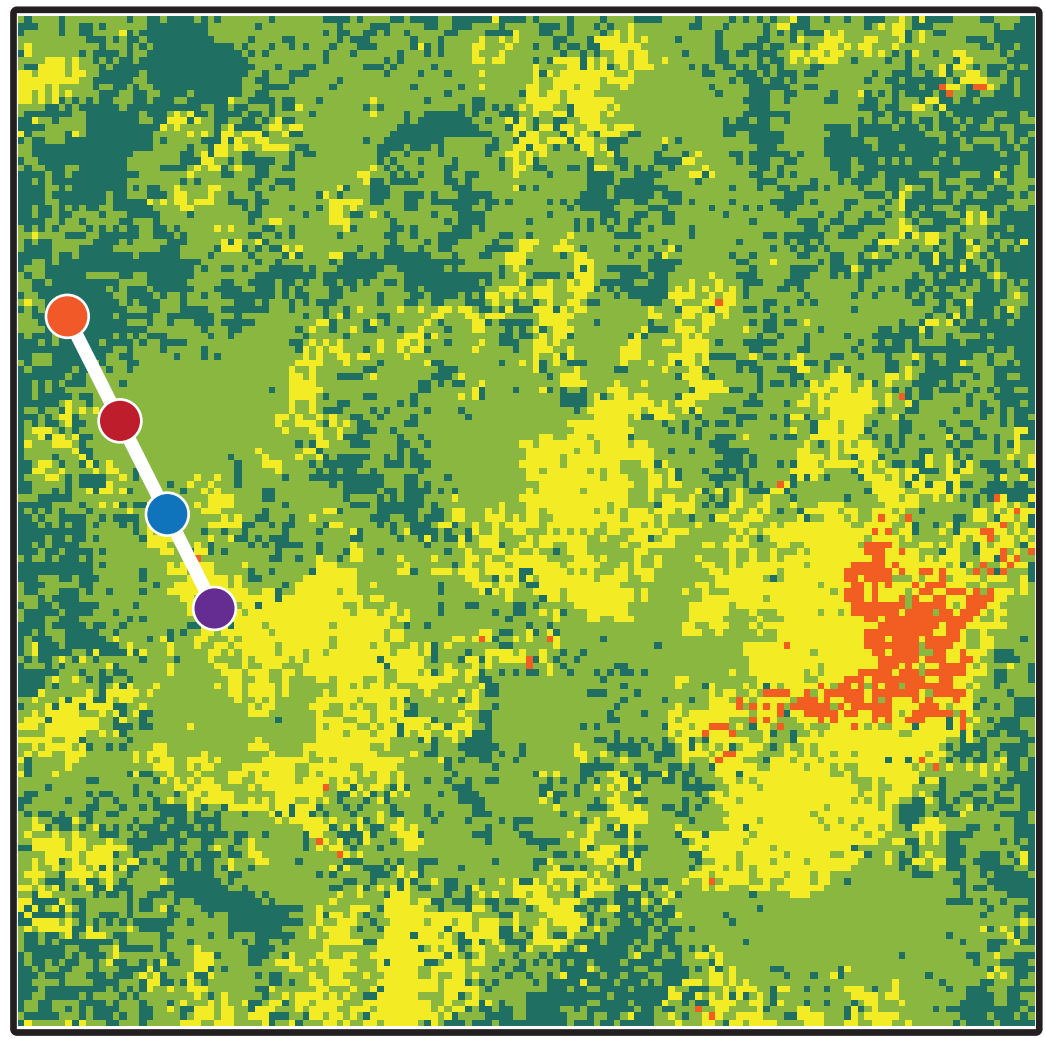
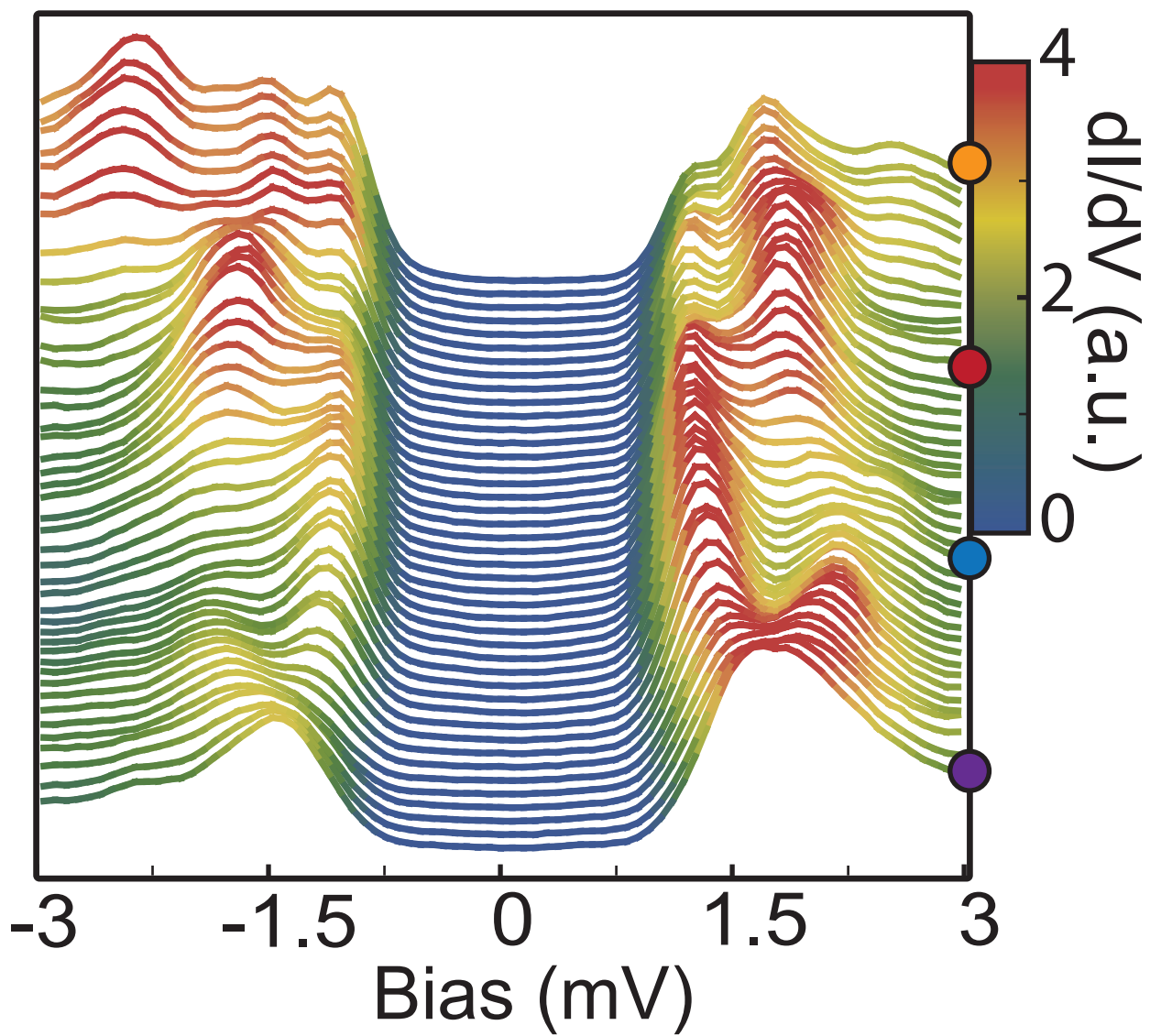
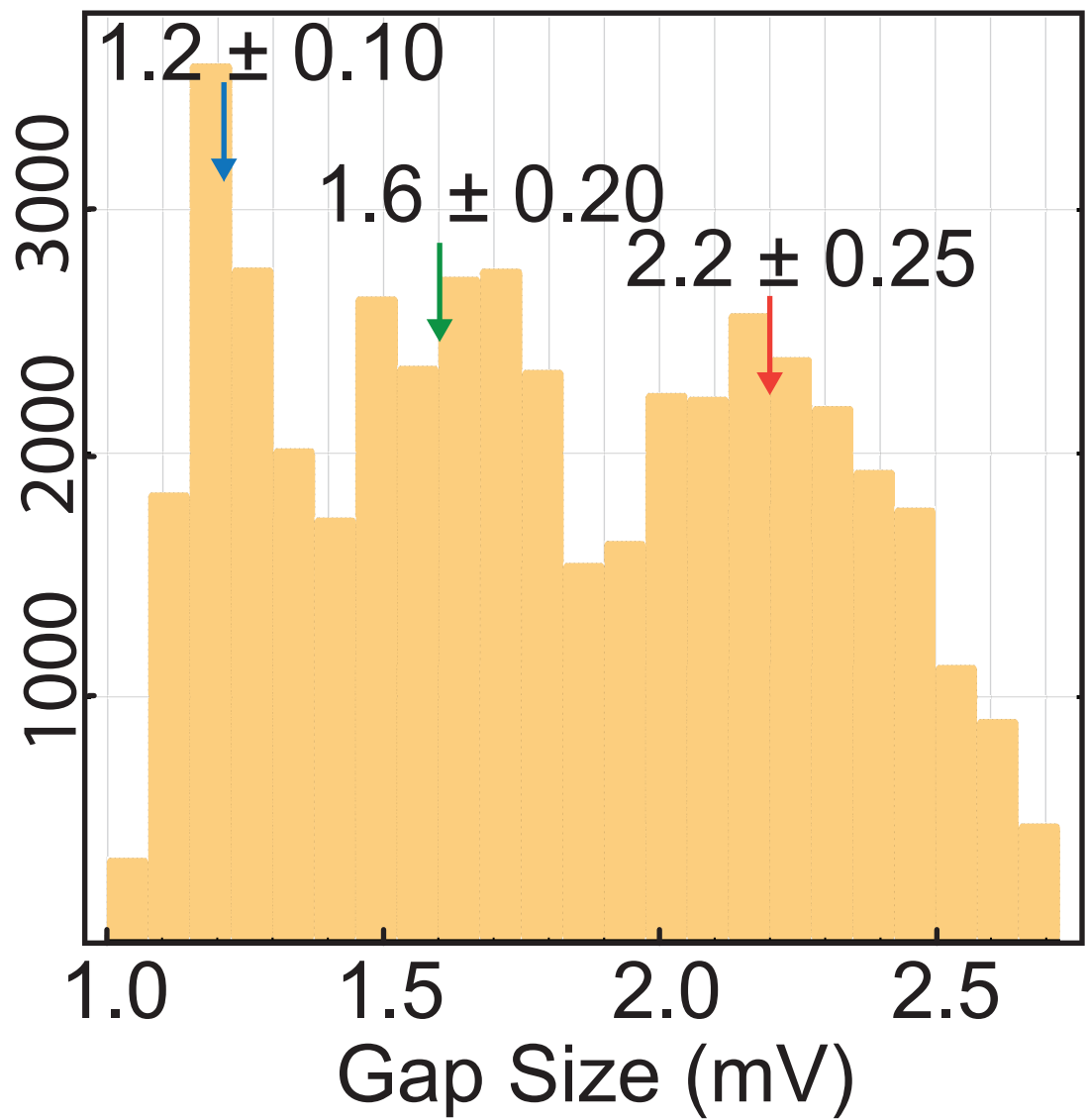

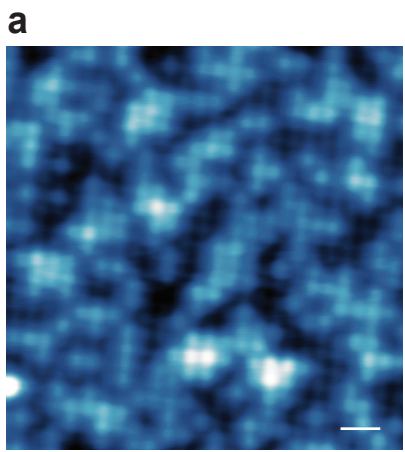
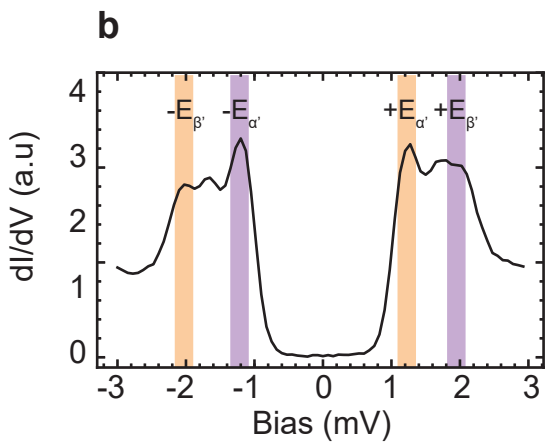
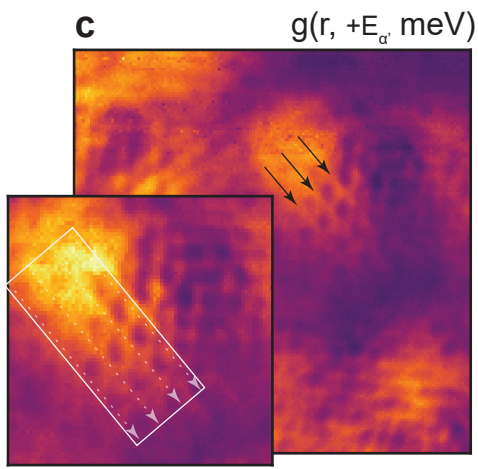
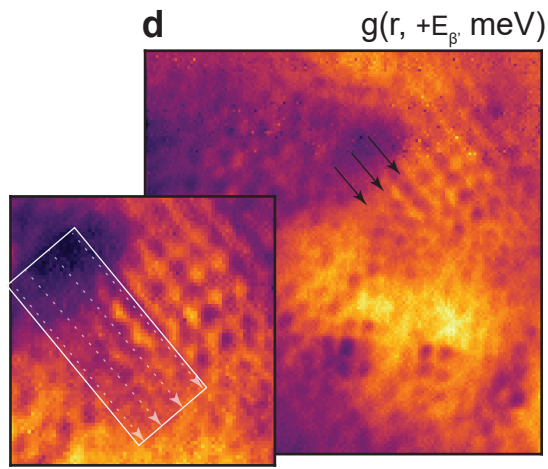
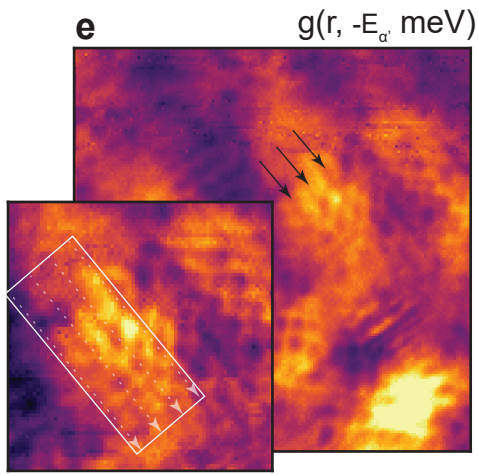
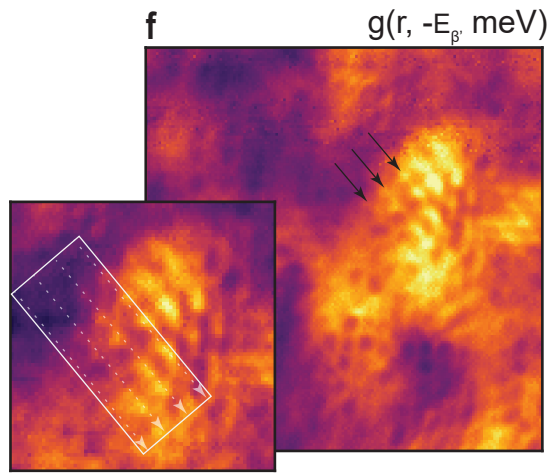
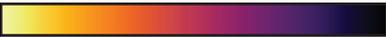

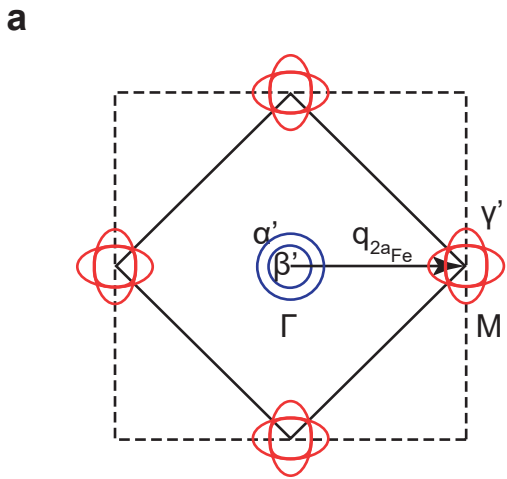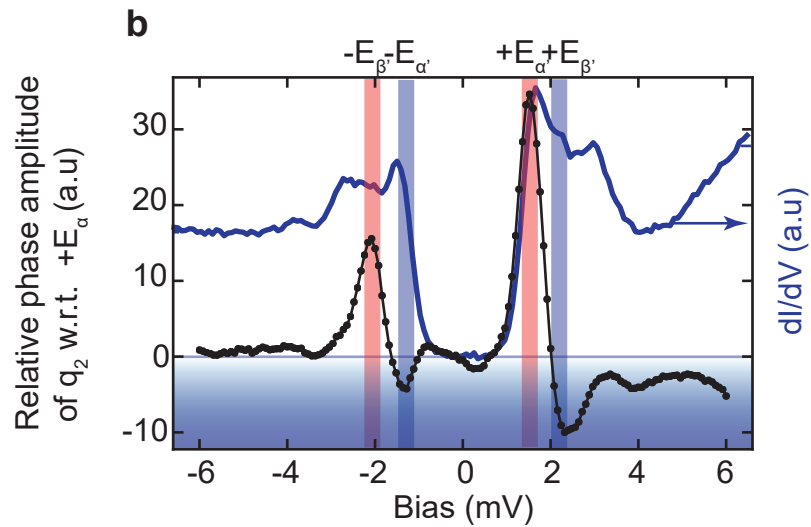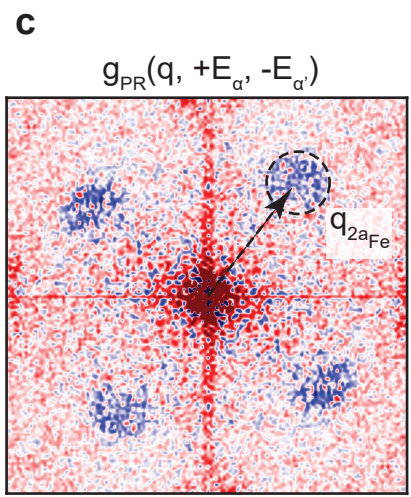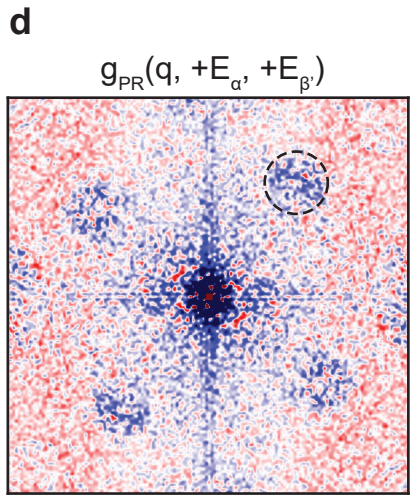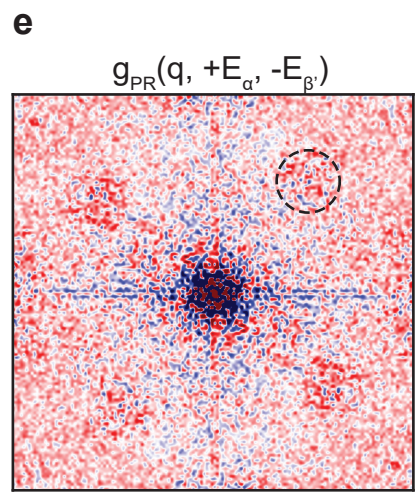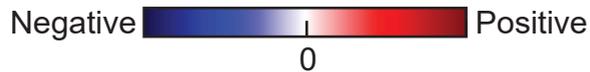